\documentclass[utf8]{frontiersSCNS}
\usepackage{url,lineno,microtype,subcaption}
\usepackage[onehalfspacing]{setspace}

\def\keyFont{\fontsize{8}{11}\helveticabold }
\def\firstAuthorLast{D'Ammando}
\def\Authors{D'Ammando Filippo\,$^{1,2*}$, on behalf of the Fermi Large Area Telescope Collaboration}
\def\Address{$^{1}$ Dipartimento di Fisica e Astronomia, Universit\'a di Bologna, Via P. Gobetti 93/2, 40129, Bologna, Italy \\
$^{2}$INAF, Istituto di Radioastronomia, Via P. Gobetti 101, 40129, Bologna, Italy}

\def\corrAuthor{D'Ammando Filippo}
\def\corrEmail{dammando@ira.inaf.it}

\begin{document}
\onecolumn
\firstpage{1}

\title[Jet Physics of Accreting SMBH in the Era of {\em Fermi}]{Jet Physics of Accreting Super-Massive Black Holes in the Era of the {\em Fermi Gamma-ray Space Telescope}} 

\author[\firstAuthorLast ]{\Authors}
\address{}
\correspondance{} 
\extraAuth{}

\maketitle

\begin{abstract}
The {\em Fermi  Gamma-ray  Space  Telescope} with its main instrument on-board, the Large Area Telescope (LAT), opened a new era in the study of
high-energy emission from Active Galactic Nuclei (AGN). When combined with contemporaneous ground- and space-based  observations, {\em Fermi}-LAT
achieves its full capability to characterize the jet structure and the emission mechanisms at work in radio-loud AGN with different black hole mass
and accretion rate, from flat spectrum radio quasars to narrow-line Seyfert 1
(NLSy1) galaxies. 

Here, I discuss important findings regarding the blazar population included in
the third LAT catalog of AGN and the $\gamma$-ray emitting NLSy1. Moreover,
the detection of blazars at redshift beyond three in $\gamma$ rays allows us
to constrain the growth and evolution of heavy black holes over cosmic time,
suggesting that the radio-loud phase may be important for a fast black hole
growth in the early Universe. Finally, results on extragalactic objects from
the third catalog of hard LAT sources are presented.  
 
\tiny\keyFont{ \section{Keywords:} active galactic nuclei, $\gamma$-ray emission, super-massive black hole, narrow-line Seyfert 1 galaxy, blazar, relativistic jet, accretion process, cosmological evolution} 
\end{abstract}

\section{Introduction}

Relativistic jets are one of the most powerful manifestations of the release
of energy related to the super-massive black hole (SMBH) at the center of
active galactic nuclei (AGN). In about 10\% of AGN, termed radio-loud AGN, the accretion disc is at
the base of a bipolar outflow of relativistic plasma, which may extend well
beyond the host galaxy, forming the spectacular lobes of plasma visible in the
radio band. The jet emission is observed across the entire electromagnetic
spectrum, from radio to $\gamma$ rays. When the jet axis is closely aligned
with our line of sight, the rest-frame radiation is strongly amplified due to
the Doppler boosting with a large fraction of the output observed at higher energies, and giving rise to the blazar phenomenon. Blazars are traditionally
divided into flat spectrum radio quasars (FSRQ) and BL Lac objects, based on
the presence or not, respectively, of broad emission lines (i.e. Equivalent Width $>$
5 \AA) in their optical and UV spectrum \citep[e.g.,][]{stickel91}. Recently,
a new classification based on the luminosity of the broad line
region (BLR) in Eddington luminosity was proposed by \citet{ghisellini11}: sources with
L$_{BLR}$/L$_{Edd}$ higher or lower than 5$\times$10$^{-4}$ being classified
as FSRQ or BL Lacs, respectively, in agreement with a transition of the
accretion regime from efficient to inefficient between these classes. The
spectral energy distribution (SED) of blazars are characterized by two bumps with
the lower energy peak occurring in the IR/optical band
in the FSRQ and at UV/X-rays in the BL Lacs. This first peak is univocally interpreted as synchrotron radiation from
highly relativistic electrons in a jet. The SED higher energy peak, observed at MeV--GeV energies in the FSRQ and at GeV--TeV energies in the BL Lacs, is commonly interpreted as inverse Compton scattering of seed photons, internal or external to the jet,
by the same relativistic electrons \citep[e.g.,][]{ulrich97}. However, other models involving hadronic processes have been proposed \citep[e.g.,][]{boettcher13}. \citet{fossati98} proposed that the SEDs of blazars form a spectral sequence, with the position of the two peaks governed by the observed bolometric luminosity: blazars with lower luminosities have the peaks at higher energies. This was theoretically interpreted by
\citet{ghisellini98} in terms of different radiative cooling suffered by the electrons emitting at the two peaks. 

Radio galaxies are viewed at larger angles than blazars, with less severe
boosting effects and classified as Fanaroff-Riley type I (FR I) and type II (FR II) based on their radio power and morphology \citep{fanaroff74}. Following
the Unified model of AGN proposed by \citet{urry95}, FR I and FR II radio
galaxies are the non-aligned (to the observer viewing angle) parent
populations of BL Lac and FSRQ, respectively. The discovery by {\em Fermi}-LAT
of variable $\gamma$-ray emission from a few radio-loud narrow-line Seyfert 1
galaxies (NLSy1) suggested this as the third class of AGN with a relativistic jet \citep{abdo09}.

The {\em Fermi  Gamma-ray  Space  Telescope} with its main instrument on-board, the Large Area Telescope (LAT), opened a new era in the study of
high-energy emission from AGN. The $\gamma$-ray sky observed by {\em Fermi}-LAT is dominated by AGN that are $\sim$60\% of the sources
included in the Third {\em Fermi} LAT source catalog \citep[3FGL;][]{acero15}. Apart from a handful of starburst galaxies, for which the $\gamma$-ray emission originates from the interaction of cosmic rays with gas and interstellar radiation fields \citep{ackermann12}, almost all the extragalactic sources are associated with radio-loud AGN.

Important findings regarding the blazar population included in the 3FGL and the NLSy1 are discussed in Section \ref{3LAC} and \ref{NLSy1}, respectively. The high-redshift blazars detected in $\gamma$ rays are discussed in Section \ref{highredshift}, while the third catalog of hard LAT sources is presented in Section \ref{3FHL}. Concluding remarks are reported in Section \ref{concluding}.

\section{The Third LAT AGN Catalogue}\label{3LAC}

Several $\gamma$-ray source catalogs, both general and for specific class of
objects (i.e. AGN, pulsars, supernova remnants, pulsar wind nebulae, gamma-ray
bursts) have already been produced by the LAT Collaboration. The 3FGL was used
as a starting point for producing a catalog of AGN only: the Third LAT AGN
catalog \citep[3LAC;][]{ackermann15}. The 3FGL includes 3033 sources detected
after four years of operation with a Test Statistic\footnote{The Test Statistic is defined as TS = 2$\times$(log$L_1$ - log$L_0$), where $L$ is the likelihood of the data given the model with ($L_1$) or without ($L_0$) a point source at the position of the target \citep[e.g.][]{mattox96}} greater than 25,
corresponding to a significance $>$ 4$\sigma$; 2192 sources are detected at Galactic latitude $\mid$b$\mid$ $>$ 10$^{\circ}$. 1563 sources (71\% of the 3FGL objects at $\mid$b$\mid$ $>$ 10$^{\circ}$) are associated at high-confidence with 1591 AGN (28 objects have two possible
associations), which constitute, together with the low-latitude AGN, the 3LAC sample. Most of the high-latitude objects (98\%) are blazars. The 3LAC includes radio-loud AGN of different types: 467 FSRQ, 632 BL Lacs, 460 BCU (Blazar with Uncertain Classification) and 32 non-blazar AGN. Among the non-blazar AGN there are 12 FR I, 3 FR II, 8 steep spectrum radio sources and 5 NLSy1.   

Associations for 182 low-latitude ($\mid$b$\mid$ $<$ 10$^{\circ}$) AGN are reported in the 3LAC: 24 FSRQ, 30 BL Lacs, 125 BCU and 3 non-blazar AGN. Extrapolating the number of high-latitude sources to low-latitude and assuming the same sensitivity, $\sim$340 sources would have been expected. The discrepancy between expected and associated source numbers is likely due to a higher Galactic diffuse emission background and a higher incompleteness of the catalogs of counterparts at low-latitude. 

On average, sources with high $\gamma$-ray luminosity (mostly FSRQ) are found
to have softer spectra than sources with low $\gamma$-ray luminosity  (mostly BL
Lacs), in agreement with the ``blazar sequence'' \citep{ghisellini98} and the
``blazar divide'' \citep{ghisellini09}, with the exception of a few outliers
(high-synchrotron-peaked-FSRQ and high-luminosity high-synchrotron-peaked BL
Lacs). A strong anti-correlation between the synchrotron peak position
($\nu_{peak}$) and the spectral index ($\Gamma_{\gamma}$) is observed for FSRQ
and BL Lacs. A similar trend is noticed for BCU supporting the idea that BCU
with low $\nu_{peak}$ and high $\Gamma_{\gamma}$ are likely FSRQ, while BCU
with high $\nu_{peak}$ and low $\Gamma_{\gamma}$ are likely BL Lacs. 
 Based on the 3LAC results, the blazar sequence was recently revised by \citet{ghisellini17}: FSRQ display approximately the same SED as the luminosity increases,
 following a sequence only in Compton dominance (i.e. the ratio of the Compton to synchrotron peak luminosities) and in the X–ray slope, while
 BL Lacs become `redder' (i.e. the peak frequencies becomes smaller) when more luminous. Moreover, a correlation between the
 jet power and the accretion power has been found in blazars, with the
 jet power dominating over the accretion disc luminosity by a factor of 10 and somewhat
 larger than the entire gravitational power \citep{ghisellini14}. 

At the time of writing, 70 AGN have been detected at TeV energies and listed in the TeVCat\footnote{http://tevcat.uchicago.edu/}. All these sources are present in the 3LAC except for HESS J1943$+$213. A large fraction ($>$ 75\%) of blazars detected by {\em Swift}-BAT in hard X-rays have also been detected by {\em Fermi}-LAT.

The blazars detected by {\em Fermi}-LAT in $\gamma$ rays after 4 years of operation represent a sizeable fraction of the entire population of known blazars as listed in the BZCAT \citep{massaro09}. The overall LAT-detected fraction is 24\% (409/1707)
for FSRQ, 44\% (543/1221) for BL Lacs and 27\% (59/221) for BCU. No strong
differences in the radio, optical and X-ray flux distributions are observed
between $\gamma$-ray detected and non-detected blazars in the BZCAT, suggesting that
all blazars could eventually shine in $\gamma$ rays at LAT-detection
level. More BL Lacs than FSRQ have been detected by {\em Fermi}-LAT so
far. This may be related to a longer duty cycle of FSRQ with respect to BL
Lacs in $\gamma$ rays.

\section{Narrow-line Seyfert 1 galaxies}\label{NLSy1}

The discovery of variable $\gamma$-ray emission from a few NLSy1 confirmed the
presence of relativistic jets in these objects. In addition to the 5 objects
reported in the 3LAC, {\em Fermi}-LAT has recently detected $\gamma$ rays from
other 3 new NLSy1: FBQS J1644$+$2619 \citep{dammando15}, B3 1441$+$476 and
NVSS J124634$+$476 \citep{dammando16}. Luminosity, variability, and spectral
properties of these NLSy1 in $\gamma$ rays indicate a blazar-like behaviour
\citep[e.g.,][]{dammando16}. Apparent superluminal jet components were detected in SBS 0846+513 \citep{dammando12}, PMN J0948$+$0022, and 1H
0323$+$342 \citep{lister16}, supporting the presence of relativistic jets in
this class of objects. 

The detection of relativistic jets in a class of AGN thought to be hosted in
spiral galaxies with a BH mass typically of 10$^{6}$-10$^{7}$ M$_{\odot}$
\citep[e.g.,][]{deo06}, challenges the theoretical scenarios of jet formation \citep[e.g.,][]{boettcher02}, suggesting two possible interpretations: either relativistic jets in NLSy1 are produced by a different mechanism or the BH mass in NLSy1 is largely underestimated. 

In the last years it has been claimed that the BH mass of NLSy1 maybe underestimated due either to the effect of radiation pressure from ionizing photons on BLR \citep{marconi08} or to projection effects \citep{baldi16}. By considering these effects, NLSy1 have BH masses of 10$^{8}$--10$^{9}$ M$_\odot$, in agreement with the values estimated by modelling the optical-UV part of their spectra with a Shakura and Sunyaev disc spectrum \citep[e.g.,][]{calderone13}. This may solve the problem of the minimum BH mass predicted in different theoretical scenarios of relativistic jet formation, but leaves open the host galaxy issue. 

Spiral galaxies are usually formed by minor mergers, with BH masses typically ranging between 10$^{6}$--10$^{7}$ M$_\odot$ \citep[e.g.,][]{woo02}, so it would not be clear how powerful relativistic jets could form in these galaxies. It is worth mentioning that the morphological classification has been done mainly for radio-quiet nearby NLSy1. Among the NLSy1 detected by {\em Fermi}-LAT up to now, the morphology of the host galaxy has been investigated only for 1H 0323$+$342, PKS 2004$-$447, and FBQS J1644$+$2619. Observations of 1H 0323$+$342 with the Hubble Space Telescope and the Nordic Optical Telescope revealed a structure that may be interpreted either as a one-armed spiral galaxy \citep{zhou07} or as a circumnuclear ring produced by a recent merger \citep{anton08, leon14}. A pseudo-bulge morphology of the host galaxy of the NLSy1 PKS 2004$-$447 and FBQS J1644$+$2619 have been claimed by \citet{kotilainen16} and \citet{olguin17}, respectively, but no conclusive results have been obtained so far. Hence, it is crucial to determine the type of galaxy hosting $\gamma$-ray emitting NLSy1 and their BH mass. 

For this reason near-infrared observations in $J$ band of FBQS J1644$+$2619
were performed using the Canarias Infrared Camera Experiment (CIRCE) at the
Gran Telescopio Canaries. The 2D surface brightness profile of the source is
modelled up to 5 arcsec by the combination of a nuclear component, associated
with the AGN contribution, and a bulge component with a S\'ersic index $n$ =
3.7, indicative of an elliptical galaxy. The structural parameters of the host
are consistent with the correlations of effective radius and surface
brightness against absolute magnitude measured for elliptical galaxies. From
the infrared bulge luminosity, a BH mass of (2.1 $\pm$ 0.2)$\times$10$^{8}$
M$_{\odot}$ was estimated \citep{dammando17}. All these pieces of evidence
strongly indicate that the relativistic jet in the NLSy1 FBQS J1644$+$2619 is produced by a massive BH in an elliptical galaxy, as expected for radio-loud AGN.

\begin{figure}[h!]
\begin{center}
\includegraphics[width=10cm]{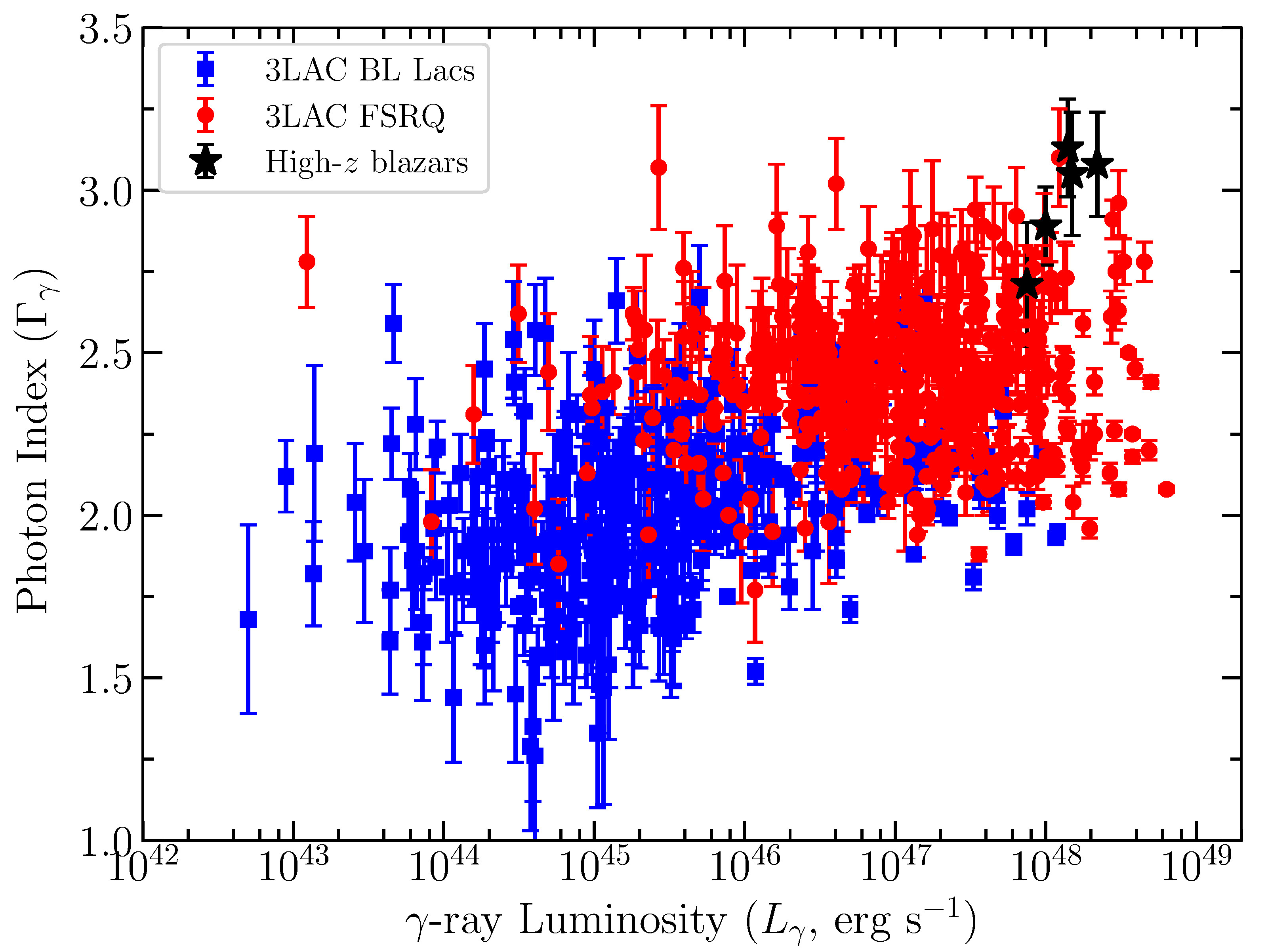}
\end{center}
\caption{Photon index vs. $\gamma$-ray luminosity plane for 3LAC and high-redshift blazars at $z$ $>$ 3.1 detected in $\gamma$ rays. Adapted from \citet{ackermann17}}\label{fig1}
\end{figure}

\section{High-redshift blazars}\label{highredshift}

High-redshift blazars are the most powerful radio-loud AGN in the Universe and
are bright targets in hard X-rays, representing a significant fraction of the
extragalactic hard X-ray sky. However, they are not commonly detected in
$\gamma$ rays. In fact, high-redshift ($z > $ 2) blazars represent $<$10\% of the AGN population observed by {\em Fermi}-LAT so far. Flaring activity in the $\gamma$-ray energy range from these sources is even more uncommon, with only fourteen FSRQ at $z$ $>$ 2 detected during a $\gamma$-ray flare.

In the 3LAC there are 64 objects at $z$ $>$ 2 ($\sim$3.7 per cent of the $\gamma$-ray sources associated with AGN) at $z$ $>$ 2, and only 2 at $z$ $>$ 3: PKS 0537$-$286 ($z$ = 3.104) and TXS 0800$+$618 ($z$ = 3.033). On the other hand, 13 blazars at $z$ $>$ 3 have been detected in hard X-rays by {\em Swift}-BAT, {\em INTEGRAL}-IBIS, and {\em NuSTAR} so far. Hard X-ray observations are more suitable for detecting blazars at $z$ $>$ 3, and this is mainly due to a spectral bias. In fact, the inverse Compton peak of high-redshift blazars is shifted towards lower energies as the bolometric luminosity increases. Only 10 sources at $z$ $>$ 2 are in both the 3LAC and the {\em Swift}-BAT 70-month catalog \citep{baumgartner13}. All
high-redshift blazars listed in both 3LAC and {\em Swift}-BAT catalogues have an average $\gamma$-ray luminosity L$_{\gamma}$ $>$ 2$\times$10$^{48}$ erg s$^{-1}$, indicating that only the most luminous blazars have been detected by both instruments. Furthermore, only blazars with an X-ray photon index $\Gamma_{\rm\,X}$ $<$1.6 have been detected in $\gamma$ rays, while no dependence on the X-ray spectral luminosity seems to be present \citep{dammando15b}.
 
As said above, high-redshift blazars at $z >$ 3.1 are missing in the {\em Fermi}
catalogs. These objects typically have large bolometric luminosities
(L$_{bol}$ $>$ 10$^{48}$ erg s$^{-1}$) and harbor extremely massive BH
(M$_{BH}$ $\sim$ 10$^{9}$ M$_{\odot}$). The new Pass 8 data set, with an improved event-level analysis, substantially enhances the
sensitivity of the LAT, in particular at lower energies, increasing the
capability of the LAT to detect sources with soft spectra like the
high-redshift blazars. By analysing 92 months of Pass 8 data between 60 MeV
and 300 GeV of a large sample of radio-loud quasars, 5 new $\gamma$-ray emitting blazars at $z$ $>$ 3.1 have
been detected with high significance. Among them, NVSS J151002+570243 ($z$ =
4.31) is now the most distant $\gamma$-ray emitting blazar so far. All the blazars discovered show steep $\gamma$-ray spectra
($\Gamma_{\gamma}$ $>$ 2.5), indicating an IC peak at MeV energies. These five
sources lie in the region of high $\gamma$-ray luminosities (L$_{\gamma}$ $>$
10$^{47}$ erg s$^{-1}$) and soft photon indices (Fig. \ref{fig1}) typical of
powerful blazars \citep{ackermann17}. Among the 5 new high-redshift blazars
there are (at least) two with redshift between 3 and 4 with a M$_{BH}$ $>$
10$^{9}$ M$_{\odot}$, implying the presence of 2$\times$2$\Gamma^{2}$
(i.e. 675, adopting $\Gamma$ = 13) similar objects but with a misaligned jet
in the same range of redshift. This changes the estimate of the space density
of very massive BH hosted in jetted sources to 68$^{+36}_{-24}$ Gpc$^{-3}$. As a consequence at $z \sim$ 4 we should have a similar number of SMBH hosted in radio-loud and radio-quiet sources and, given their strong evolution, above that redshift most massive BH might be hosted in radio-loud AGN. This suggests that the radio-loud phase can be a key ingredient for a rapid BH growth in the early Universe.

\section{The Third Catalog of Hard Fermi-LAT Sources}\label{3FHL}

\begin{figure}[h!]
\begin{center}
\includegraphics[width=10cm]{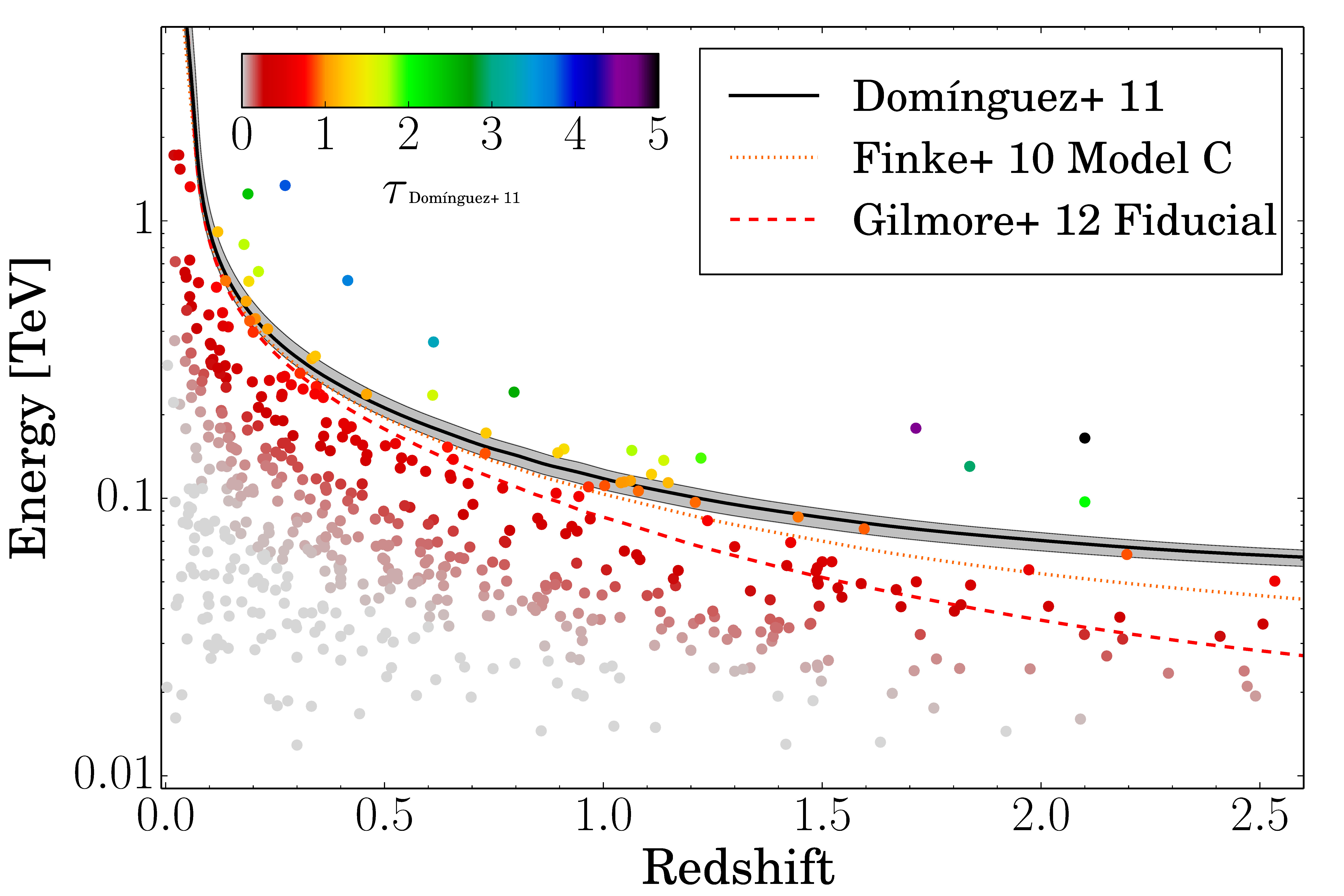}
\end{center}
\caption{The highest-energy photons vs. redshift for 3FHL blazars, color coded by the optical depth calculated from the model presented by \citet{dominguez11}. The cosmic $\gamma$-ray horizon based on different EBL models is shown. Adapted from \citet{ajello17}}\label{fig2}
\end{figure}

In addition to the {\em Fermi}-LAT catalogs with the standard low-energy threshold of 100
MeV, three hard source catalogs have been released: the First {\em Fermi}-LAT
Catalog of Sources above 10 GeV \citep[1FHL;][]{ackermann13}, based on the first three
years of data analyzed in the 10--500 GeV energy range, the Second Catalog of Hard
{\em Fermi}-LAT Sources \citep[2FHL;][]{ackermann16}, based on 80 months of
data analyzed in the 50 GeV--2 TeV energy range, and the Third Catalog of Hard
{\em Fermi}-LAT Sources \citep[3FHL;][]{ajello17}, based on 7 years of data in the
10 GeV--2 TeV energy range. The 3FHL contains 1556 objects and takes advantage
of the improvement provided by Pass 8 by using the PSF-type event
classification. The 3FHL includes 214 new $\gamma$-ray sources never appeared
in previous {\em Fermi} catalogs. Three of these 214 have been detected with
the Imaging Atmospheric Cherenkov Telescopes (IACT). The vast majority of
detected sources (79\%) are associated with extragalactic counterparts at
other wavelengths, including 16 sources located at high-redshift ($z$ $>$ 2):
11 FSRQ, 3 BL Lac, and 2 BCU. BL Lacs are the most numerous extragalactic
population (61\%) followed by BCU (23\%) and FSRQ (14\%). Only 72 of the 3FHL 
extragalactic sources have been already detected by current IACT. In this context, the 3FHL is a resource for planning observations of the current
(MAGIC, VERITAS, H.E.S.S.) and future (Cherenkov Telescope Array) IACT observatories.
Interestingly, a few highest-energy photons from distant blazars included in the 3FHL catalogue are
in the region around and beyond the cosmic $\gamma$-ray horizon (i.e. the energy at which the cosmic optical depth $\tau$ = 1, see e.g. \citet{dominguez13}) as shown in Fig. \ref{fig2}. These photons provide important constraints on extragalactic background light models as they may also help in the understanding of $\gamma$-ray propagation over cosmological distances.

\section{Concluding Remarks}\label{concluding}

{\em Fermi}-LAT has been performing the first all-sky survey in $\gamma$ rays, gathering well-sampled, continuous light curves for hundreds
of AGN and compiling source catalogs for different energy ranges and time
periods. These observations constitute important resources to the astronomical
community for a better understanding of the jet physics, cosmological
evolution, and accretion processes of SMBH. In fact, the {\em Fermi}-LAT
observations, in conjunction with the multi-frequency data collected from
radio to VHE, are key to revealing the nature of jet physics in different
classes of AGN, including particle acceleration, environmental effects, and
interaction processes. In addition, Pass 8 LAT data have increased the
sensitivity for hard-spectrum sources, which are important targets for
ground-based VHE telescopes including the planned Cherenkov Telescope
Array. Moreover, the extended energy range reached by the Pass 8 data opens
new opportunities for the study of blazars at high redshifts. Over the next
years the {\em Fermi} satellite will provide a fundamental contribution in time domain astronomy and multi-messenger/multi-wavelength studies.

\section*{Conflict of Interest Statement}

The author declares that the research was conducted in the absence of any commercial or financial relationships that could be construed as a potential conflict of interest.

\section*{Author Contributions}

The author confirms being the sole contributor of this work and approved it for publication.

\section*{Acknowledgments}

The \textit{Fermi}-LAT Collaboration acknowledges support for LAT development, operation and data analysis from NASA and DOE (United States), CEA/Irfu and IN2P3/CNRS (France), ASI and INFN (Italy), MEXT, KEK, and JAXA (Japan), and the K.A.~Wallenberg Foundation, the Swedish Research Council and the National Space Board (Sweden). Science analysis support in the operations phase from INAF (Italy) and CNES (France) is also gratefully acknowledged. This work performed in part under DOE Contract DE-AC02-76SF00515.

\end{document}